\documentclass[a4paper,fleqn,usenatbib]{mnras}

\usepackage[T1]{fontenc}
\usepackage{ae,aecompl}

\usepackage{graphicx}	
\usepackage{amsmath}	
\usepackage{amssymb}	

\title[Secular Resonance of Iapetus]{Secular Resonance Between Iapetus and the Giant Planets}

\author[{\'C}uk et al.]{
Matija {\'C}uk,$^{1}$\thanks{E-mail: mcuk@seti.org (MC)}
Luke Dones,$^{2}$
David Nesvorn{\'y}$^{2}$
and Kevin J. Walsh$^{2}$
\\
$^{1}$SETI Institute, 189 North Bernardo Avenue, Suite 200, Mountain View, CA 94043, USA\\
$^{2}$Southwest Research Institute, 1050 Walnut Street, Suite 400, Boulder, CO 80302, USA
}

\date{Accepted XXX. Received YYY; in original form ZZZ}

\pubyear{2018}

\begin{document}
\label{firstpage}
\pagerange{\pageref{firstpage}--\pageref{lastpage}}
\maketitle

\begin{abstract}
Using numerical integrations, we find that the orbital eccentricity of Saturn's moon Iapetus undergoes prominent multi-Myr oscillations. We identify the responsible resonant argument to be $\varpi-\varpi_{g5}+\Omega-\Omega_{eq}$, with the terms being the longitudes of pericenter of Iapetus and planetary secular mode $g_5$, Iapetus's longitude of the node and Saturn's equinox. We find that this argument currently (on a $10^7$~yr timescale) appears to librate with a very large amplitude. On longer timescales, the behavior of this resonant angle is strongly dependent on the resonant interaction between Saturn's spin axis and the planetary mode $f_8$, with long-term secular resonance being possible if Saturn's equinox is librating relative to the node of the $f_8$ eigenmode. We present analytical estimates of the dependence of the resonant argument on the orbital elements of Iapetus. We find that this Iapetus-$g_5$ secular resonance could have been established only after the passage of Iapetus through the 5:1 mean-motion resonance with Titan, possibly in the last Gyr. Using numerical simulations, we show that the capture into the secular resonace appears to be a low-probability event. While the Iapetus-$g_5$ secular resonance can potentially help us put new constraints on the past dynamics of the Saturnian system, uncertainties in both the spin axis dynamics of Saturn and the tidal evolution rate of Titan make it impossible to make any firm conclusions about the resonance's longevity and origin.
\end{abstract}


\section{Introduction}

Iapetus is the third-largest moon of Saturn, as well as the major moon that is the most distant from the planet. Iapetus is notable for its albedo dichotomy \citep{bur95, por05}, oblate shape \citep{tho07, cas11}, and equatorial ridge \citep{lev11, dom12, sti18}, but here we will restrict ourselves to studying its orbital motion. Like other regular satellites, Iapetus has a relatively low orbital eccentricity ($e_I=0.028$), but it also has a substantial orbital inclination ($i_I=8^{\circ}$ with respect to its Laplace plane\footnote{The instantaneous Laplace plane can be defined for every perturbed orbit as plane normal to the vector around which the orbit normal is precessing.}), the origin of which has been a long-standing problem \citep{war81, nes14}. As the solar perturbations on Iapetus's orbit are comparable to those arising from Saturn's oblateness and the inner moons (chiefly Titan), the Laplace plane of Iapetus is significantly tilted to Saturn's equator($i_L=14^{\circ}$). As Iapetus's orbit precesses around its Laplace plane, the instantaneous inclination of Iapetus to Saturn's equator varies approximately over a $5^{\circ}-21^{\circ}$ range over Iapetus's nodal precession period of about 3400~yr. 

Iapetus's inclination contradicts the established opinion that Iapetus and other regular satellites formed from a flat disk surrounding Saturn. Any disk consisting of gas and/or small particles that is inclined to the local Laplace plane would be subject to differential nodal precession at different distances. Through collisions and other dissipative mechanisms, the disk would soon settle into the local Laplace plane. A satellite that forms from such a disk should have no inclination at all. Therefore, if Iapetus formed in orbit around Saturn (as suggested by its prograde, low-eccentricity orbit), some dynamical process had to impart inclination to Iapetus after its formation. \citet{war81} suggested that Iapetus's inclination could have been generated through rapid gas disk dissipation. If the circumplanetary disk could disappear in a time comparable to or shorter than the 3400-year nodal precession period of Iapetus, the resulting change in the Laplace plane could induce a substantial free inclination. However, it is not clear that the circumplanetary disk would disappear on such a short timescale \citep{mar11}.

Another potential source of Iapetus's inclination would be close encounters between Saturn and ice giants during planetary migration \citep{tho99, tsi05}. If these encounters were to operate as a classic random-walk process, they would excite a distant satellite's eccentricity more than its inclination \citep{pah15}. However, \citet{nes14} found that in a significant number of planetary flybys they simulated, the inclination of Iapetus was excited by several degrees while its eccentricity stayed well below 0.01. This behavior was associated with distant encounters ($r>0.1$~AU), and the inclination excitation was apparently driven by secular torques from highly-inclined passing ice-giants, which had little effect on the eccentricity. Such distant encounters between Saturn and the ice giants were also found to be capable of capturing the existing irregular satellites of Saturn \citep{nes07a, nes14a}.

Recently, there has been some reconsideration of the dynamical history of the Saturnian system, prompted by observations of much faster than expected tidal evolution \citep{lai12, lai17}. While in the classical picture \citep[e.g.][]{md99} Iapetus does not take part in any resonances with other satellites, faster tidal evolution would make Titan and Iapetus cross their mutual 5:1 mean-motion resonance in the past. This crossing should have happened about 500 Myr ago if we assume a uniform tidal quality factor $Q=1500-2000$ for all satellites \citep{cuk13}, or could have happened at a very different epoch if the tidal evolution of Saturn's moons is driven by resonant modes inside the planet \citep{ful16}. Since this paper deals with the relatively recent past (a few hundred Myr), we will mostly assume that Titan's orbital evolution is driven by Saturn's constant tidal quality factor $Q=1500$ and tidal Love number $k_2$ (tidal evolution of Iapetus is negligible in this model).  

\section{Current Dynamics of Iapetus with a Fixed-Obliquity Saturn}

We start our study by importing position and velocity vectors for Iapetus, Titan and the four giant planets (with the epoch of January 1, 2000) from the Jet Propulsion Laboratory's HORIZONS ephemeris system\footnote{https://ssd.jpl.nasa.gov/?horizons accessed on January 24, 2013}. We use these vectors as initial conditions in simulations using numerical integrators derived from {\sc simpl}, which was previously employed by \citet{cuk16}. Briefly, {\sc simpl} is a mixed-variable, symplectic integrator based on an algorithm of \citet{cha02} that simultaneously integrates the orbits of the planets and satellites of one of the planets. The basic version of {\sc simpl} includes all mutual perturbations (except the satellites' effects on planets), as well as the parent planet's oblateness, tidal torques on satellites and additional migration forces (to account for ring or disk torques, when necessary). One important limitation of {\sc simpl} is that the planet's spin axis is stationary and not affected by any of the torques that would act on it in the real system (this includes both precession-inducing gravitational torques and tidal dissipation with the planet). In the case of Saturn, this approximation is justified when studying the relatively fast dynamics of the inner satellites \citep{cuk16}, as their orbital precession periods are on the order of years and decades, while the precession period of Saturn's spin axis is longer than 1 Myr \citep{fre17}. Even when dealing with Titan and Iapetus, precession periods are still shorter than $10^4$ yr, seemingly making Saturn's pole precession irrelevant. However, when studying longer-period dynamics, precession and other motions of Saturn's spin axis will need to be taken into account, as detailed below.

\begin{figure}
	\includegraphics[width=\columnwidth]{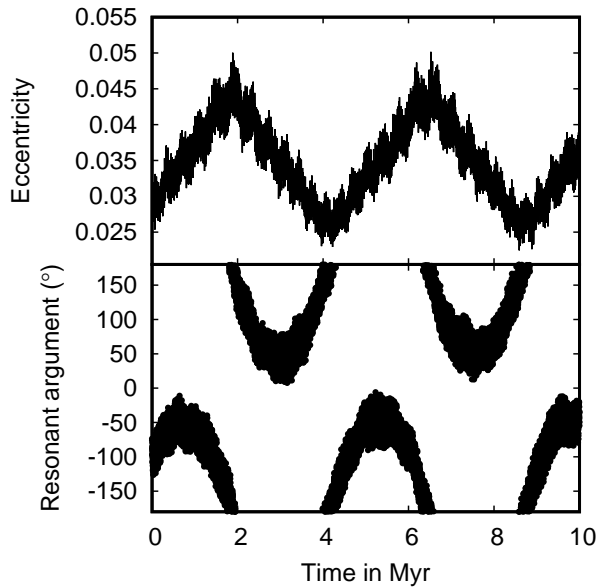}
    \caption{Top: Eccentricity of Iapetus during a 10 Myr integration of Iapetus's orbit using {\sc psimpl}. Bottom: Evolution of the resonant argument $\varpi-\varpi_J+\Omega-\Omega_{eq}$ in the same simulation.}
    \label{psim2}
\end{figure}

Our first and simplest modification of {\sc simpl} so we can study the dynamics of Iapetus's orbit over Myr timescales is the introduction of uniform precession of Saturn's spin axis around the invariable plane. The version of {\sc simpl} modified in this manner is designated {\sc psimpl}, with a "{\sc p}" signifying precession. Figure \ref{psim2} (top panel) shows the evolution of Iapetus's eccentricity over 10~Myr integrated using {\sc psimpl}, assuming Saturn's axial precession period to be 1.96~Myr. In this integration we included the full orbital dynamics of the four giant planets, as well as Titan and Iapetus. We ignored Hyperion and included the satellites interior to Titan into Saturn's $J_2$ obliquity term. A periodic variation with a period of about 4~Myr is clearly present in Fig. \ref{psim2}, with the variation comparable to the average eccentricity of Iapetus. This variation is clearly caused by a very slow-changing resonant (or near-resonant) argument, and its very long period compared to the 3400-year apsidal precession period of Iapetus (which is the conjugate of angular momentum and must be present in a eccentricity-affecting term) suggests a near-canceling of two similar precession terms. \citet{cuk16} found a somewhat similar resonance involving the sum of apsidal and nodal precessions of the inner moons. The near-identical precession rates (with opposite signs) of the apsidal and nodal precession for Tethys and Dione produce very slow-changing secular terms, with a rate of change more than two orders of magnitude slower than the basic precession frequencies \citep{cuk16}. This inspired us to investigate terms including the angle $\varpi+\Omega$, which has a period of about $3 \times 10^5$~yr. This term is close to secular resonance with the $g_5$ mode of planetary eccentricities (i.e. the ``slow'' or ``aligned'' mode of Jupiter and Saturn). In order to satisfy the D'Alembert rules for the arguments of the disturbing function \citep{md99}, an additional very slowly evolving node-type angle is necessary; we opted for the longitude of Saturn's equinox (with respect to the invariable plane), as it determines the orientation of Iapetus's Laplace plane. The evolution of the resulting resonant argument $\varpi-\varpi_J+\Omega-\Omega_{eq}$ is plotted in the bottom panel of Fig. \ref{psim2}, where $\varpi$ and $\varpi_J$ are the longitudes of pericenter of Iapetus and Jupiter\footnote{Here and elsewhere in this paper we used $\varpi_J$ as a directly-observable proxy for the orientation of the $g_5$ eccentricity vector, as Jupiter's eccentricity is dominated by the $g_5$ mode.}, while $\Omega$ and $\Omega_{eq}$ are the longitudes of Iapetus's ascending node and Saturn's vernal equinox. Fig. \ref{psim2} clearly indicates that a term with this argument is responsible for the variations in Iapetus's eccentricity, and that the resonant argument appears to librate with a large amplitude over the next 10 Myr. 

While the secular resonances usually evolve on precession timescales, Iapetus-$g_5$ secular resonance described here has a more slowly evolving argument involving $\varpi+\Omega$, i.e. the sum of the apsidal and nodal precession rates of the same body, which are usually opposite and approximately equal for regular satellites. The only other currently known examples of a similar resonant argument among regular satellites are the Pallene-Mimas secular resonance found by \citet{cal10}, and the past Tethys-Dione secular resonance proposed by \citet{cuk16}, and in both cases the secular resonance is caused by proximity to a mean-motion resonance (MMR). Resonances including combinations of $\varpi+\Omega$ are also found among asteroids, where they are referred to as the $z_1$ and $z_2$ secular resonaces \citep{mil92, mil94}. 

Among regular satellites with orbital precession dominated by the planet's oblateness, the angles $\varpi + \Omega$ precess at rates that decrease monotonically with distance from the planet, and an additional perturbation (such as a nearby MMR) is needed to make these angles for two moons enter a resonance.  The unique dynamics of Iapetus, which is at the transition between oblateness-dominated and solar perturbation-dominated orbits, allows for the observed secular resonance in the absence of any MMRs. Iapetus has about the slowest orbital precession that is possible for a Saturnian satellite, which then places the rate of change of its $\varpi+\Omega$ angle right in the parameter space occupied by planetary secular frequencies (in this case $g_5$). We will address the relevant terms affecting the precession of the angle $\varpi+\Omega$ in more detail in Section 4.

\begin{figure}
	\includegraphics[width=\columnwidth]{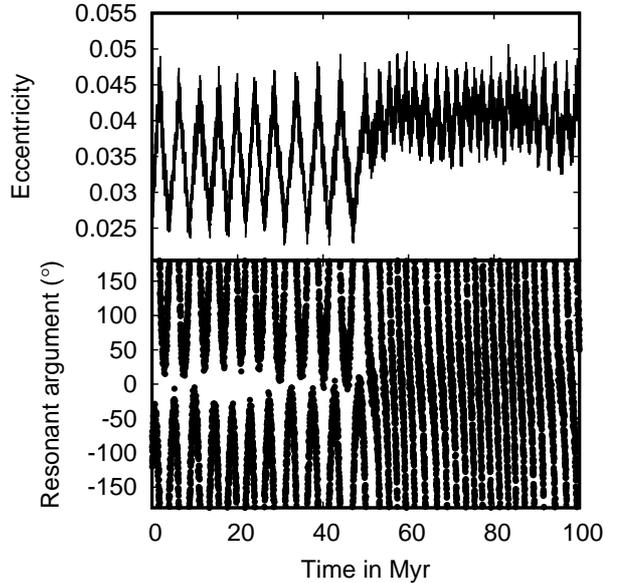}
    \caption{Top: Eccentricity of Iapetus during a 100 Myr integration of Iapetus's orbit using {\sc psimpl}. Bottom: Evolution of the resonant argument $\varpi_I-\varpi_J+\Omega_I-\Omega_{eq}$ in the same simulation. Here we used a constant precession of Saturn's spin axis around the invariable plane with a 1.96 Myr period, as in Fig. \ref{psim2}.}
    \label{psim3}
\end{figure}

\begin{figure}
	\includegraphics[width=\columnwidth]{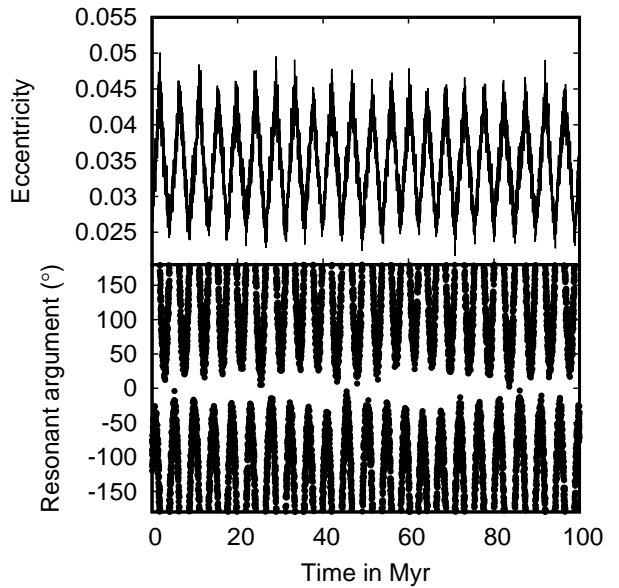}
    \caption{Top: Eccentricity of Iapetus during a 100 Myr integration of Iapetus's orbit using {\sc psimpl}. Bottom: Evolution of the resonant argument $\varpi_I-\varpi_J+\Omega_I-\Omega_{eq}$ in the same simulation. Here we used a constant precession of Saturn's spin axis around the invariable plane with a 1.91 Myr period, equal to the period of the secular mode $f_8$ \citep{md99}.}
\label{psim4}
\end{figure}

In Fig. \ref{psim2}, the resonant argument of the Iapetus-$g_5$ secular resonance librates for 10 Myr under the above stated assumptions of Saturn's pole precession. To study longer-term stability of this resonance, we extended the simulation to 100 Myr, and the results are plotted in Fig. \ref{psim3}. About halfway through the integration in Fig. \ref{psim3}, the Iapetus-$g_5$ secular resonance breaks and the argument enters circulation (bottom), while the eccentricity now oscillates with only about half of the previous amplitude. Before deciding that the secular resonance is ephemeral, we need to consider the limitations of our model. Apart from the assumption of constant-rate, constant-obliquity precession of Saturn's pole built into {\sc psimpl}, we also had to select a precession rate for Saturn. The rate we chose (-0.66~arcsec yr$^{-1}$, with a 1.96~Myr period) is based on the observational results of \citet{fre17} for the current precession rate of Saturn's pole (-0.45~arcsec yr$^{-1}$), which had to be converted into the long-term average rate. \citet{war04} \citep[using the moon precession models of][]{vie92} find that the current precession rate of Saturn's pole should be about 68\% percent of the long-term rate due to the 700-year cycle of Titan's orbital precession. Therefore we used that value to adjust the results of \citet{fre17}, obtaining the rate of -0.66~arcsec yr$^{-1}$. Given the approximate way we combined the results of these authors, it is very likely that the evolution plotted in Fig. \ref{psim3} does not reflect the real dynamics of the system.

Another way to estimate the long-term precession rate of Saturn's pole is to assume that it is locked in a spin-orbit secular resonance with the node associated with the $f_8$ secular mode of the Solar System \citep{war04, ham04}. The precession rate of the secular mode $f_8$ is -0.69~arcsec~yr$^{-1}$, equivalent to a period of 1.91~Myr \citep{md99, las11, vok15, zee17}. Figure \ref{psim4} shows the 100~Myr evolution of Iapetus's eccentricity and Iapetus-$g_5$ secular resonant argument using {\sc psimpl} and assuming the $f_8$ precession rate for Saturn's pole. In this case, libration in the Iapetus-$g_5$ resonance is preserved over 100~Myr, with the current periodic oscillations in eccentricity persisting throughout the simulation. This demonstrates the sensitivity of the Iapetus-$g_5$ secular resonance to the precessional dynamics of Saturn's spin axis, and shows the need for a more sophisticated model of Saturn's precessional motion, which we will address in the next section.

\section{Current Dynamics of Iapetus with a Variable Obliquity of Saturn}

In order to model the full spin dynamics of Saturn, we needed to modify {\sc simpl} further to include the realistic response of Saturn's spin axis to solar, satellite and planetary torques. Since Saturn's precessional period is much slower than any of the periods studied here, we are justified in using an azimuthally symmetric, oblate model of Saturn, despite known azimuthal asymmetries \citep{elm17}. Similarly, the large distances between interacting bodies involved here (Titan is the closest perturber) justify restricting ourselves to the $J_2$ moment of Saturn (which also includes Rhea and interior satellites). We decided to use the same approach as \citet{cuk16b} did for Earth in their integrations of the Earth-Moon system. In every timestep, Saturn's spin axis suffered a ``kick'' \citep[cf.][]{vok15}:
\begin{equation}
d{\bf {\hat n}} = {\Sigma \ 3 m_i J_2 ( {\bf r}_i \times {\bf{\hat n}} ) ({\bf r}_i \cdotp {\bf {\hat n}}) dt \over \alpha R^2 \omega_R r_i^5}
 \end{equation}
where ${\bf{\hat n}}$ is the spin axis unit vector, $m_i$ is the mass of the perturber (in units of AU$^3$~yr$^{-2}$), $J_2$ is the usual oblateness moment (including effective oblateness due to satellites interior to Titan), ${\bf r}_i$ is the radius-vector of the perturber w.r.t. Saturn, $dt$ is the timestep, and $\alpha$, $R$ and $\omega_R$ are Saturn's dimensionless moment of inertia, radius and spin rate, respectively. While the orbits of Titan and Iapetus were affected by the oblateness of Saturn (effectively the reverse of the above torque, but calculated independently), we ignored the back-reaction of Saturn's spin on heliocentric orbits. We interwove this kick with the other perturbations in the usual ``leapfrog'' manner. While this is the simplest possible implementation of Saturn's spin dynamics in a fully numerical integrator, we find that there are no discernible errors over the 100s of Myr we studied (which are only hundreds of Saturn's precession periods). We term the version of {\sc simpl} with a freely precessing planet {\sc ssimpl}, with the extra ``{\sc s}'' standing for ``spin''.
\begin{figure}
	\includegraphics[width=\columnwidth]{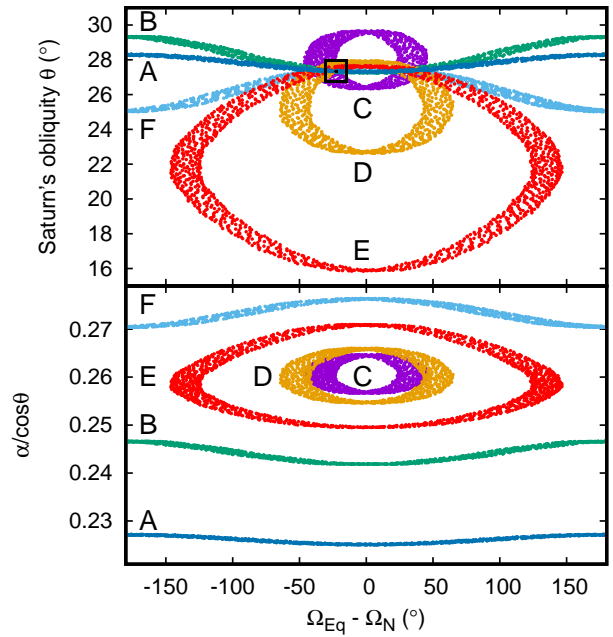}
    \caption{(Top) Evolution of Saturn's obliquity as a function of the spin-orbit resonant argument $\Omega_{eq}-\Omega_N$ and over the next 100~Myr, obtained using {\sc ssimpl} and six different values for Saturn's moment of inertia; see Table \ref{table} for details. The square symbol plots the initial conditions (i.e., the current state). (Bottom) The same integrations, now with the ratio of the moment of inertia $\alpha$ and the cosine of obliquity (as a measure of precession rate) plotted on the $y$-axis. Both $\Omega_{eq}$ and $\Omega_N$ were determined with respect to the invariable plane of the Solar System.}
\label{eye}
\end{figure}

In {\sc ssimpl}, as it fully integrates the precessional dynamics, the only adjustable parameter is Saturn's principal moment of inertia $\alpha$, which then determines Saturn's angular momentum. In reality, $\alpha$ is convolved with the differential rotation of Saturn to produce angular momentum, but here we will use a constant rotation rate of 5211.3~rad~yr$^{-1}$, which corresponds to a period of 10.569~h. The value of $\alpha$ is not known directly, and the observations of Saturn's pole precession are the most promising way of measuring it. Therefore we integrated Saturn's pole precession (and the associated dynamics of Iapetus) for six different values of $\alpha$, which we refer to as cases A-F (Table \ref{table}). Figure \ref{eye} shows some of the solutions (A, B, F) circulating and some (C-E) librating, meaning that the pole of Saturn is in secular resonance with Neptune's longitude of the node.\footnote{Here and throughout we used $\Omega_N$ as a directly-observable proxy for the phase of the secular eigenmode $f_8$. The presence of other modes in Neptune's inclination vector leads to some smearing in the $x$-direction of the curves plotted in Fig. \ref{eye}.} 

\begin{table}
	\centering
	\caption{Parameters for the six different integrations plotted in Figs. \ref{eye} and \ref{prec}. The units for precession rates are arcsec~yr$^{-1}$, and all the values are negative. The third column lists the long-term precession rates (with respect to the invariable plane) fitted over $\simeq 1$~Myr, while the fourth column lists the average precession rates (with respect to the ecliptic) for the 1975-2015 period. The values in square brackets are not fits to integrations but estimates (assuming 61.5\% of the third column).}
	\label{table}
	\begin{tabular}{cccc} 
		\hline
		Case & $\alpha$ & prec. rate & obs. rate \\
		\hline
		A & 0.2 & 0.799 & [0.49]\\
		B & 0.215 & 0.744 & [0.46]\\
		C & 0.23 & 0.6955 & 0.427\\
		D & 0.235 & 0.681 & 0.42\\
		E & 0.24 & 0.667 & 0.41\\
		F & 0.245 & 0.6535 & [0.40]\\
		\hline
	\end{tabular}
\end{table}

\begin{figure}
	\includegraphics[width=\columnwidth]{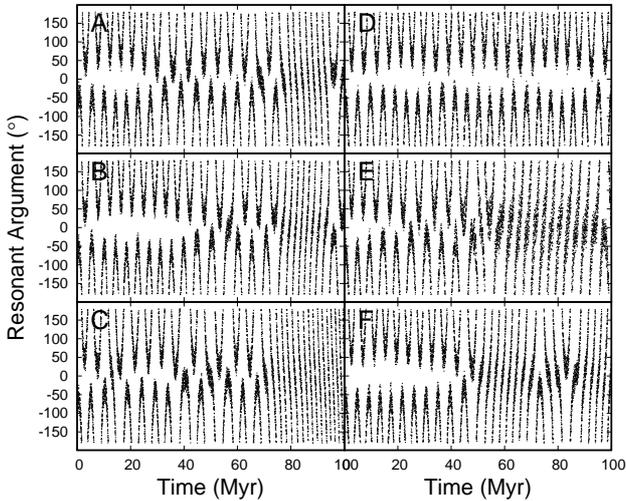}
    \caption{Evolution of the Iapetus-$g_5$ secular resonant argument over the next 100~Myr, obtained using {\sc ssimpl} and six different values for Saturn's moment of inertia. See Table \ref{table} for details.}
\label{prec}
\end{figure}

Figure \ref{prec} plots the evolution of the Iapetus-$g_5$ secular argument over each of the six 100~Myr simulations. While Iapetus is initially in resonance, it remains in the resonance for the whole of 100~Myr only in one of the six cases: case D, in which Saturn's pole is librating in the spin-orbit resonance, with the present obliquity being close to the maximum one. However, it is not certain that the difference between the different solutions is systematic, and not stochastic, and to answer that question we would need to run many more computationally intensive simulations. However, if we could identify the correct solution for Saturn's pole precession from observations, we could greatly constrain the problem and we should be able to predict the future of the Iapetus-$g_5$ secular resonance with more confidence.

\begin{figure}
\includegraphics[width=\columnwidth]{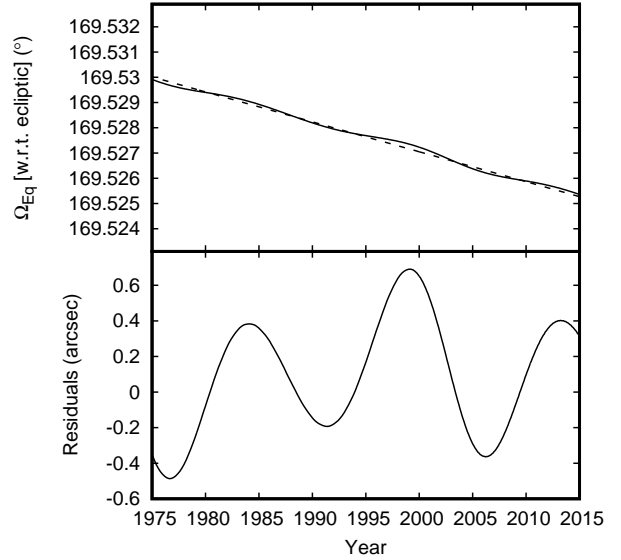}
\caption{(Top panel) The precession of Saturn's longitude of equinox (with respect to the ecliptic) in the 1975-2015 period in our simulation C. A short-term average precession rate of $-0.427$~arcsec~yr$^{-1}$ is plotted with a dashed line. (Bottom panel) Residuals between the Saturn's longitude of the equinox and the linear fit using the rate of $-0.427$~arcsec~yr$^{-1}$. The most prominent periodic feature is associated with the $2 \lambda_S - 2 \Omega_{eq}$ term.}
\label{obs}
\end{figure}

Determining where the current system is among the six simulations plotted in Fig. \ref{eye} is non-trivial. \citet{fre17} have measured the rate of Saturn's pole precession sinve the Voyager encounter to be $-0.451 \pm 0.014$~arcsec~yr$^{-1}$. This rate cannot be compared directly to the long-term precession rate, which we fit to our simulations over $\simeq 1$~Myr and list in the third column of Table \ref{table} for every simulation (1~Myr is longer than most of the periodic terms but shorter than the libration in the spin-orbit resonance). In order to be able to compare the theory and observation more directly, we also computed current observable precession rates for Saturn's pole (for the years 1975-2015, and relative to the J2000 ecliptic, rather than the invariable plane we use in our long-term fits). The short-term fits are listed for cases C, D, and E (which have a  librating pole of Saturn) in Table \ref{table}. We find that the current precession rate is between $-0.41$ and $- 0.427$~arcsec~yr$^{-1}$ for the three librating cases, which can be compared to the value reported by \citet{fre17}. Formally, our case C is within 2~$\sigma$ of the observed value, and \citet{fre17} state that their formal errors may underestimate the true uncertainties. Figure \ref{obs} plots the short-term variation in the longitude of Saturn's equinox for 1975-2015 in simulation C (top panel), and the same results with the average rate of $-0.41$~arcsec~yr$^{-1}$ subtracted (bottom panel). A strong periodic feature proportional to $\sin(2 \lambda_S - 2 \Omega_{eq})$ (where $\lambda_S$ is Saturn's mean longitude), with an amplitude of $\simeq 0.4$~arcsec~yr$^{-1}$, can be seen in the bottom panel. We speculate that this periodic term has lowered the precession rate in \citet{fre17} Fit \#1, which is based on Cassini data for 2004-2010, as well as for pole position variations in their Figure 12. In any case, it is clear that a linear fit is not sufficient to fit Saturn's pole precession to observations with high accuracy.

\begin{figure}
	\includegraphics[width=\columnwidth]{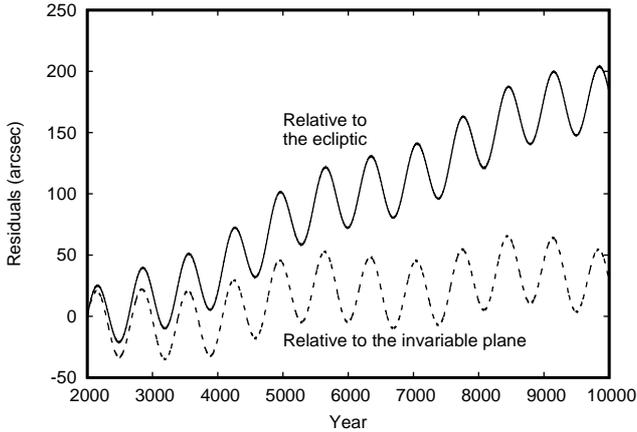}
    \caption{The residuals between the longitude of Saturn's equinox and the long-term average precession rate -0.6955~arcsec~yr$^{-1}$ in our simulation C. The solid line plots the longitude of the equinox measured relative to the ecliptic J2000, while the dashed line plots one with respect to the invariable plane (the latter definition was used to calculate the long-term average precession rate). The periodic terms due to Titan (700~yr period) and Iapetus (3500~yr period) are visible, and the slight upward trend in the dashed line is due to a 50,000~yr $\Omega_{eq}-\Omega_S$ term.}
\label{obs2}
\end{figure}

A related question is why our simulations suggest that the current precession rate is about 61.5\% of the long term one, while \citet{war04} \citep[based on the model of][]{vie92} obtained 68\%. Figure \ref{obs2} plots the evolution of Saturn's longitude of equinox in simulation C over the next 8000~yr, once the long-term average rate of 0.6955~arcsec~yr$^{-1}$ has been removed. First of all, the precession rate is different when measured relative to the ecliptic (as observers do) and the invariable plane of the Solar System (which is used in theoretical calculations). Apart from Titan's main 700~yr nutation period, we can also see a smaller periodic term associated with Iapetus (with a period of about 3500~yr) and a secular trend associated with the $\Omega_{eq}-\Omega_S$ periodic term which has a period of about 50,000~yr. While \citet{vie92} definitely included the dynamics of Iapetus into their model, \citet{war04} took only the dominant effect of Titan into account when determining the current/mean precession ratio of 68\%. We conclude that our numerical simulations may be consistent with past analytical estimates once all periodic terms are included. We also find that the maximum allowed value for Saturn's moment of inertia from the empirical model of \citet{hel09}, $\alpha=0.226$, would still put Saturn's pole in libration within the spin-orbit resonance, due to the substantial resonance width. 

The results of this section indicate that the current state of knowledge does not allow us to predict the long-term stability of the Iapetus-$g_5$ secular resonance, or even the future behavior of Saturn's spin pole. However, it is clear that data can be consistent with Saturn's pole librating in the resonance with secular eigenmode $f_8$, and \citet{war04} and \citet{ham04} have made a strong case on theoretical grounds that this resonance is present. The phase of Saturn's pole precession close to the libration center is also indicative of the resonance. It is tempting to use the Iapetus-$g_5$ secular resonance as the constraint in Saturn's pole precession, i.e. argue that simulation D is closest to the real solution as it preserves the Iapetus-$g_5$ secular resonance. However, we cannot make such pronouncements based on six simulations, and also it is not impossible that the Iapetus-$g_5$ secular resonance is short-lived or intermittent, especially if it is less than 1~Gyr old (Section 5). We conclude that the way forward will be to compare observations of Saturn's pole position to a full numerical model of its precession, which is outside the scope of this paper.

\section{Location of the \lowercase{$g_5$} Secular Resonance}

A secular argument which includes the angle $\varpi+\Omega$ evolves very slowly, due to near-cancellation of the precession rates of the pericenter and the node. In a first-order approximation, these two precessional rates are indeed the same (with an opposite sign), so we have to look to higher order terms to identify the sources of secular frequencies.

The first effect to consider is orbital precession due to perturbations from Saturn's oblateness and all the satellites interior to Iapetus (including Titan). While the leading term oblateness-driven precession is symmetric for the pericenter and the node, the symmetry is broken for eccentric and inclined orbits \citep{dan92}:
\begin{equation}
{\dot \Omega}_2 = - {3 J_2 n \over 2 (a/R)^2 (1-e^2)^2} \cos{i}
\end{equation}
\begin{equation}
{\dot \omega}_2 = {3 J_2 n \over 2 (a/R)^2 (1-e^2)^2} ({5 \over 2} \cos^2{i} -{1 \over 2})
\end{equation}
So, assuming small-inclination orbits, the sum of Iapetus's apsidal and nodal precession is ${\dot \varpi}+ {\dot \Omega} = {\dot \omega}+ 2 {\dot \Omega}$:
\begin{equation}
{\dot \varpi}_2+ {\dot \Omega}_2 = {3 J_2 n \over 4 (a/R)^2 (1-e^2)^2} (5 \cos^2{i} - 1 - 4 \cos{i})
\end{equation}
We introduce $s=\sin{i}$, and assume that both $e$ and $i$ are small quantities (so $\cos{i}=1-s^2/2$, and we ignore $O(e^2 s^2)$, $O(s^4)$):
\begin{equation}
{\dot \varpi}_2+ {\dot \Omega}_2 = {3 J_2 n \over 4 (a/R)^2} (5 - 5 s^2 -1 - 4 + 2 s^2) = - {9 J_2 n \over 4 (a/R)^2} s^2
\label{j2}
\end{equation}
To the lowest order, the sum of apsidal and nodal precession due to planetary oblateness does not depend on eccentricity and is negative (i.e. retrograde) for orbits with non-zero inclination. For Iapetus, this term amounts to about -5.0~arcsec~yr$^{-1}$; note that 3/4 of the $J_2$ in Eq. \ref{j2} comes from Titan, and the rest mostly from Saturn's oblateness. While $J_2^2$ term is the dominant non-zero part of the sum of apsidal and nodal precession for the inner moons \citep{cuk16}, it can be ignored for Iapetus, due to its small size and dependence on distance as $(a/R)^{-4}$ \citep{md99}.

Next terms we need to consider are Titan's perturbations not included in the $J_2$ term. Since the orbits of Titan and Iapetus are relatively well-separated ($a/a_T \simeq 3$), we will restrict ourselves to terms arising from $J_4$ perturbations of Titan (here $J_4=(3/8) (m_T/M)(a_T/R)^4)$). The precession of a satellite's orbit due a $J_4$ moment is given in \citet{bro59}:
\begin{equation}
\dot{\omega}_4 =  B_4 [21 - 9 \eta^2 + (-270 + 126 \eta^2) \vartheta^2 + (385 - 189 \eta^2) \vartheta^4]
\end{equation}
\begin{equation} 
\dot{\Omega}_4 =  4 B_4 [(5 - 3 \eta^2) \vartheta (3 - 7 \vartheta^2)]
\end{equation}
Where
\begin{equation}
B_4={15 J_4 n \over 128 (a/R)^4 (1-e^2)^4}
\end{equation}
where $\vartheta=\cos{i}$ and $\eta=\sqrt{1-e^2}$. Since we are interested in the small $e$ and $i$ case, we switch from $\eta$ and $\vartheta$ to $e$ and $s$:
\begin{equation}
\dot{\omega}_4 \simeq B_4 [64 + 72 e^2 - 248 s^2]
\end{equation}
\begin{equation} 
\dot{\Omega}_4 \simeq 4 B_4 [-8 -12 e^2 +18 s^2] 
\end{equation}
The net contribution to the rate of change of the secular resonance argument is then:
\begin{equation}
\dot{\varpi}_4+\dot{\Omega}_4 = {15 J_4 n \over 16 (a/R)^4} (-3 e^2 - 13 s^2)
\end{equation}
which equates to about -1.8~arcsec~yr$^{-1}$ for Iapetus.

The next term we need to examine is one due to solar secular perturbations, commonly referred to Kozai-Lidov interaction \citep{lid62, koz62}. The expression for the precession of a satellite's orbit due to solar quadrupole-order perturbations averaged over mean motions (assuming no coupling between mean-motion and secular terms) is \citep{inn97, cuk04}:
\begin{equation}
\dot{\omega}_K = K_2 [2 (1-e^2) + 5 \sin^2{\omega}(e^2 -\sin^2{i}]
\end{equation}
\begin{equation}
\dot{\Omega}_K = - K_2 [1 + 4 e^2 -5 e^2 \cos^2{\omega}] \cos{i}
\end{equation}
where 
\begin{equation}
K_2 = {3 n_S^2 \over 4 (1-e_S^2) \sqrt{1-e^2} n}
\end{equation}
where $e_S$ and $n_S$ are Saturn's eccentricity and mean motion, respectively. Since Iapetus has very weak oscillations in $e$ and $i$ as $\omega$ precesses, we can average over $\omega$, so $\cos^2{\omega}=\sin^2{\omega}=1/2$. Therefore, assuming small $s$, we get:
\begin{eqnarray}
\dot{\omega}_K = K_2 (2 + {1 \over 2} e^2 - {5 \over 2} s^2) \\
\dot{\Omega}_K = - K_2 (1 + {3 \over 2} e^2 - {1 \over 2} s^2)
\end{eqnarray}
So, finally, the Kozai contribution to the change of secular resonance argument is:
\begin{equation}
\dot{\varpi}_K+\dot{\Omega}_K \simeq {3 n_S^2 \over 8 n} (- 5 e^2 - 3 s^2)
\end{equation}
which for Iapetus amounts to -8.0~arcsec~yr$^{-1}$.

The three secular contributions derived above amount to about -14.8~arcsec~yr$^{-1}$, which would make the secular resonance argument $\dot{\varpi} +\dot{\Omega}$ precess in the retrograde direction with a period $< 10^5$~yr, while numerical integrations show this angle in resonance with $\dot{\varpi}_J+\dot{\Omega}_{eq}$ which has the prograde precession rate of 3.6~arcsec~yr$^{-1}$. Before we lose faith in secular theory, we must remember that the Kozai-Lidov precession assumes no coupling between short-period and secular terms, which fails spectacularly when applied to apsidal precession of the Moon, as discovered by Clairaut \citep{bau97}. In reality, there is notable coupling between the apsidal precession of the satellite and the mean motion of the Sun, leading to the so-called ``evection'' term. While the averaged evection term is much more important for secular behavior of the Moon and irregular satellites which are much more strongly perturbed by the Sun, is is relevant to the Iapetus-$g_5$ secular resonance as it affects the pericenter much more strongly than the node. The leading Clairaut terms for the precession driven by evection, as well as the analogous coupling between the solar mean-motion and the nodal precession, are \citep{cuk04}:
\begin{equation}
\dot{\varpi}_C + \dot{\Omega}_C = \Bigl( {225 \over 32 }(1-s^2)+ {9 \over 32} \Bigr) {n_S^3 \over n^2} + {4071 \over 128}{n^4_S \over n^3}
\end{equation}  
Here we kept only the inclination dependence of the largest apsidal term, and ignored the $e$ and $i$ dependence for the other two. This gives us a positive precession contribution of 17.7~arcsec~yr$^{-1}$ to the rate of change of the secular resonance argument. 

For completeness, since we are accounting for non-zero average contributions of periodic terms, we must include the average effect of the octupole apsidal secular interaction between Titan and Iapetus. If we average the precessional effects of the $\varpi_T-\varpi$ term \citep{lee03} the same way \citet{cuk04} did for solar evection, we get:
\begin{equation}
\dot{\varpi}_3= {225 \over 512} {n^2 \over (\dot{\varpi}_T-\dot{\varpi})} \Bigl( {m_T \over M}\Bigr)^2 \Bigl({a_T \over a}\Bigr)^6 {e_T^2 \over e^2} 
\label{octupole}
\end{equation}
where $(\dot{\varpi}_T-\dot{\varpi})$ has a period of about 900~yr, and the net contribution of this term is 0.7~arcsec~yr$^{-1}$. This brings the total $\dot{\varpi} + \dot{\Omega}$ rate from all five contributing terms (effective $J_2$ and $J_4$, Kozai-Lidov, Clairaut and averaged-octupole) to a prograde rate of 3.6~arcsec~yr$^{-1}$, which is within 0.1~arcsec~yr$^{-1}$ of $\dot{\varpi}_J+\dot{\Omega}_{eq}$ and therefore satisfactorily explains the current secular resonance of Iapetus. 

It is interesting that the current very slow resonance exist only because several larger terms mostly cancel each other out. Since the major positive contribution to $\dot{\varpi} + \dot{\Omega}$  is from the Clairaut terms which (to the first order) do not depend on $e$ and $i$ like the retrograde terms, the fastest $\dot{\varpi} + \dot{\Omega}$ possible at Iapetus's distance from the Sun would be 17.4~arcsec~yr$^{-1}$ for circular orbits in the Laplace plane. So a resonance with the $g_6$ planetary eigenmode is not possible in the same way as the observed one with the $g_5$ eigenmode. The only other secular resonance we found for Iapetus has the argument $\varpi+\Omega-\varpi_T-\Omega_T$ which requires an almost zero $\dot{\varpi} + \dot{\Omega}$ for Iapetus, and happens at somewhat higher $e$ and/or $i$.
\begin{figure}
	\includegraphics[width=\columnwidth]{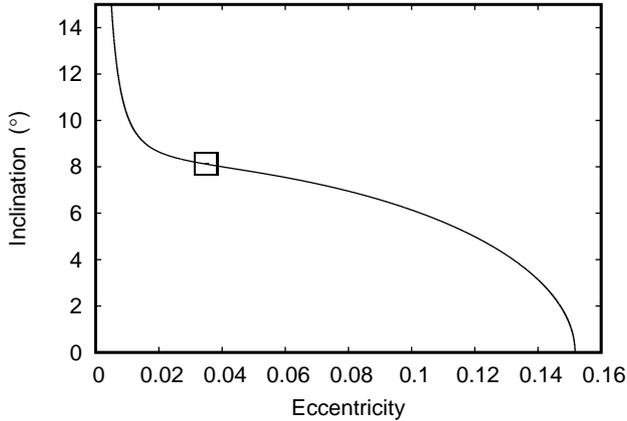}
    \caption{The location in the $e-i$ plane where $\dot{\varpi} + \dot{\Omega}=3.6$~arcsec~yr$^{-1}$ according to Eq. \ref{total}. The open square plots the approximate current average orbital elements of Iapetus. The behavior of the curve close to $e=0$ and $i=0$ is poorly constrained, as at those points the definitions of $\varpi$ and $\Omega$ become unreliable.}
\label{loc}
\end{figure}
Now we can plot where the resonance is in the $e-i$ phase space for Iapetus, assuming the current semimajor axes for Iapetus, Titan and Saturn (this level of approximation ignores the eccentricity of Saturn). Using all of the precession contributions we derived, we get:
\begin{equation}
\dot{\varpi} + \dot{\Omega} = P_0 - P_i \sin^2{i} - P_e e^2 + P_3 e^{-2}
\label{total}
\end{equation}
where
\begin{equation}
P_0= {117 \over 16} {n_S^3 \over n^2} + {4071 \over 128}{n^4_S \over n^3}
\end{equation}
\begin{equation}
P_e = {15 \over 8} \Bigl({{ 3 J_4 n \over 2 (a/R)^4 } + {n^2_S \over n}}\Bigr)
\end{equation}
\begin{equation}
P_i= {3 \over 8}\Bigl( { 6 J_2 n \over (a/R)^2} + {65 J_4 n \over 2 (a/R)^4 } + { 3 n^2_S \over n} + { 75 n^3_S \over 4 n^2}\Bigr)
\end{equation}
and $P_3$ is defined by Eq. \ref{octupole}. In Fig. \ref{loc} we plot the locations in the $e-i$ plane where $\dot{\varpi} + \dot{\Omega}=3.6$~arcsec~yr$^{-1}$ according to Eq. \ref{total}. There is a continuous line of locations in $e-i$ phase space for which the resonance is possible, along which eccentricity decreases while the inclination increases.  
 
It is tempting to observe Fig. \ref{loc} and envision past evolution (or diffusion) of Iapetus's orbit along the secular resonance. The orbit of Iapetus is more inclined and less eccentric than would be expected (on average) from excitation by planetary fly-bys \citep{nes14}, and some of the combinations of $e$ and $i$ along the resonant location would be a more likely outcomes of encounters of Saturn with the Ice Giants. Diffusion of orbits along the secular resonance is known to be a major effect in the dynamics of asteroids and meteoroids \citep{nes07}. However, the resonant perturbations must affect $e$ and $i$ equally and with the same sign, as the coefficients of $\varpi$ and $\Omega$ in the resonant argument are the same. Therefore secular resonant perturbations (chaotic or not) can only move Iapetus's orbit along a diagonal line in $e-i$ space along which both $e$ and $i$ are increasing or decreasing at the same time. However, this line is close to perpendicular to the line plotting the locations of the secular resonance, greatly reducing the potential for evolution or diffusion through the resonance. So it is likely that the secular resonance with the $g_5$ mode was established only once Iapetus reached its current orbit, and could not be responsible for its unusually high inclination and much lower eccentricity. 

\section{The Origin of the Secular Resonance}

Iapetus's semimajor axis is practically fixed, as the tidal acceleration of Iapetus due to tides on Saturn raised by Iapetus is negligible \citep{lai12}. However, Titan does migrate appreciably due to tides, which changes the Titan/Iapetus mean motion ratio and affects the secular dynamics of Iapetus. The most notable dynamical event in the history of the Titan-Iapetus pair was likely their mutual 5:1 mean-motion resonance crossing \citep{cuk13, pol17, pol18}. If we ignore the satellite tides within Titan, this crossing should have happened about 0.5~Gyr ago, assuming $Q/k_2 \simeq 5000$ for Saturn \citep{lai12}. Prior work has found that this resonance can excite the eccentricity of Iapetus from zero to the present value, while the inclination of Iapetus could not have been changed substantially. While in the majority of cases Iapetus survives this resonance crossing when we assume $e < 0.01$ for Titan, Iapetus is almost always lost if Titan had its present eccentricity of $e_T=0.029$ during MMR crossing \citep{cuk13, pol17, pol18}. This prompted \citet{cuk16} to propose that the eccentricity of Titan was recently ($\simeq$100~Myr) excited, as a side-effect of a massive instability among the inner moons. 

\begin{figure}
	\includegraphics[width=\columnwidth]{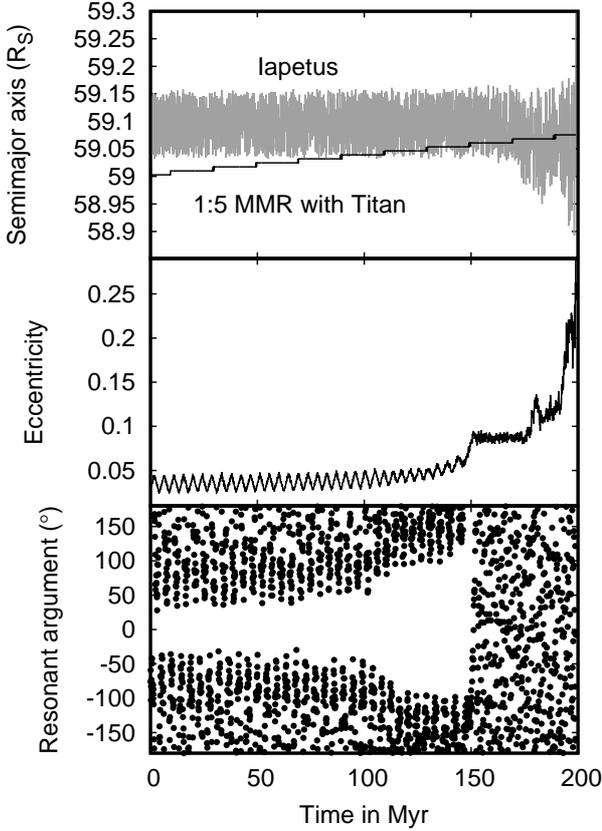}
    \caption{Evolution of Iapetus's orbit using {\sc psimpl} and assuming that Iapetus was in the $g_5$ secular resonance (with present $e$ and $i$) before the Titan-Iapetus 5:1 MMR. We used an axial precession rate of $-0.069$~arcsec~yr$^{-1}$ (same as in Fig. \ref{psim4}) and a tidal $Q/k_2=5000$ for Saturn. The secular resonance moves to higher eccentricities due to proximity of the 5:1 MMR, making Iapetus more eccentric in the process. Iapetus enters the 5:1 MMR with $e \simeq 0.1$, which usually leads to instability. Here and in Figs. \ref{sres42}-\ref{sres52}, the ``stair step" texture of Titan's semimajor axis plot is an artifact of a low-precision output.}
\label{res6}
\end{figure}

The first question to ask is whether the Iapetus-$g_5$ secular resonance can be ancient, pre-dating the Titan-Iapetus 5:1 MMR crossing. Figure \ref{res6} shows a simulation (using {\sc psimpl}, see caption for details) in which we integrated the 5:1 resonance crossing with Iapetus initially in the secular resonance. In this and other simulations we consistently get eccentricity growth for Iapetus as it approaches the 5:1 MMR. Apparently, Titan's near-resonant perturbations on Iapetus lead to a positive precession of the secular resonance argument, most likely by affecting ${\dot\varpi}$ \citep[cf. ][ who found similar secular-MMR interference for the Dione-Rhea 5:3 resonance]{cuk16}. This additional positive rate of change of the secular resonance argument must be balanced by the increase in eccentricity in order to preserve the resonance (as the $e^2$ term in Eq. \ref{total} has a negative coefficient). The secular resonance is broken when $e$ reaches about 0.1. Eccentricity is then constant until the 5:1 MMR is encountered, at which point the orbit of Iapetus becomes chaotic and is eventually destabilized. Since this is a systematic result, we conclude that Iapetus was unlikely to be in the secular resonance with the $g_5$ mode before the 1:5 MMR with Titan and that the resonance must have been established more recently.

\begin{figure}
	\includegraphics[width=\columnwidth]{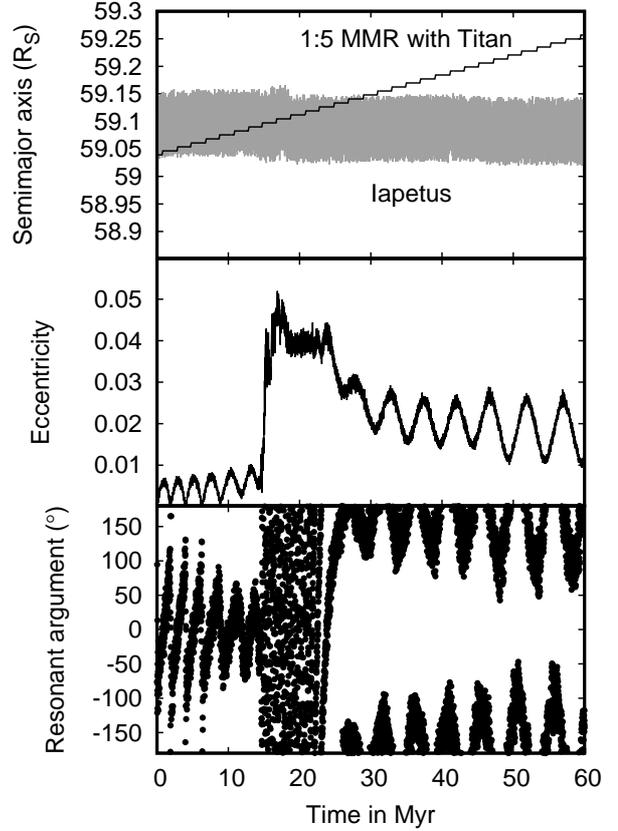}
    \caption{A simulation of the Titan-Iapetus 5:1 MMR using {\sc ssimpl}, with tidal evolution accelerated 10x. The bottom panel plots the secular resonant argument $\varpi_I-\varpi_J+\Omega_I-\Omega_{eq}$. In this run, Iapetus is captured into the secular resonance soon after crossing the 5:1 MMR. }
\label{sres42}
\end{figure}

\begin{figure}
	\includegraphics[width=\columnwidth]{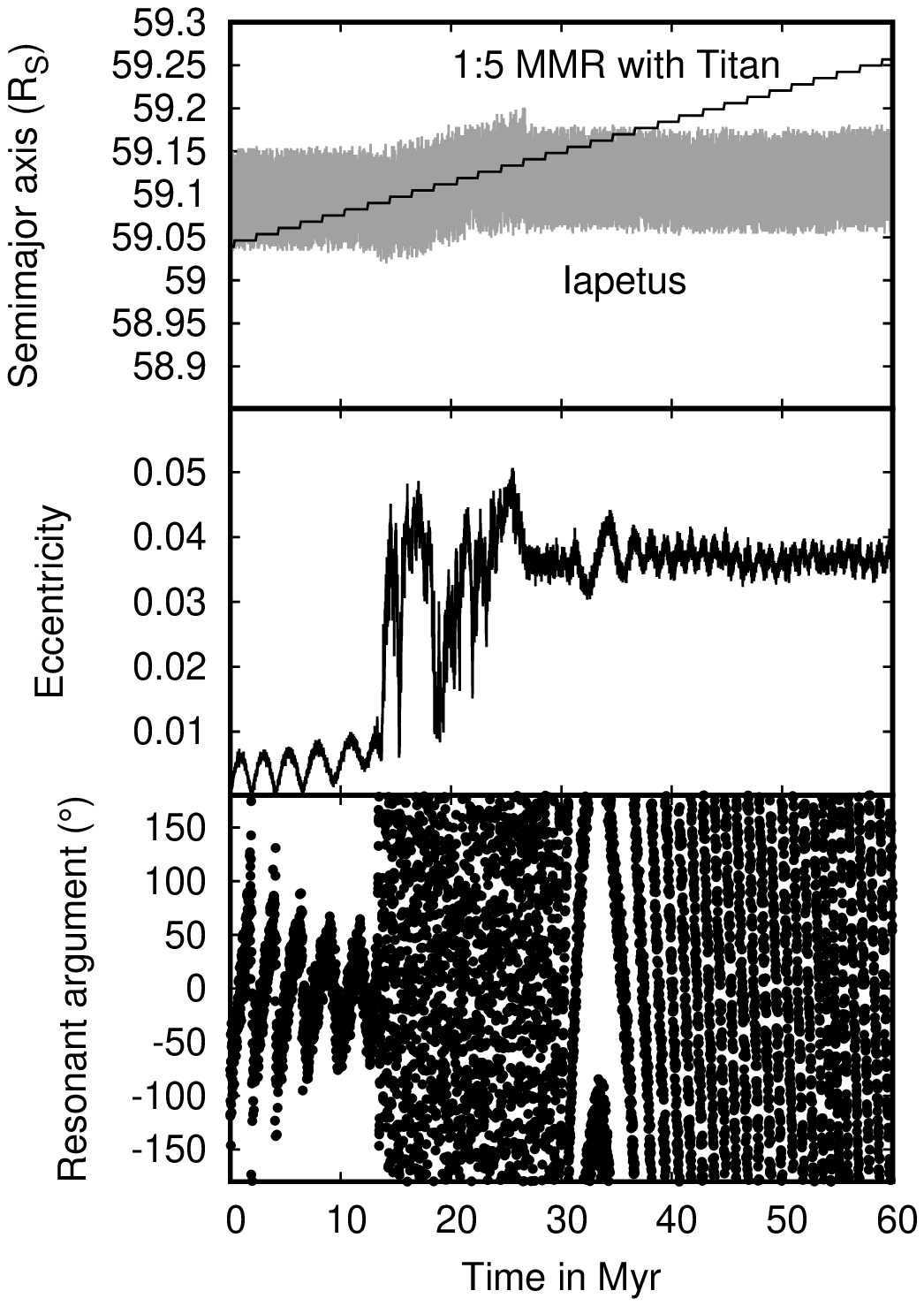}
    \caption{A simulation of the Titan-Iapetus 5:1 MMR using {\sc ssimpl}, with tidal evolution accelerated 10x. The bottom panel plots the secular resonant argument $\varpi_I-\varpi_J+\Omega_I-\Omega_{eq}$. In this run, Iapetus crosses the secular resonance at about 33~Myr without being captured.}
\label{sres34}
\end{figure}

An intriguing possibility is an evolution that is in a way reversed from that shown in Fig. \ref{res6}: Iapetus encountering the $g_5$ secular resonance as it exits the 5:1 MMR, subsequently evolving to lower eccentricities. One problem in modeling this process is the high degree of stochasticity. The outcome of the 5:1 MMR resonance is unpredictable even with low initial eccentricities for Titan and Iapetus. Sometimes Iapetus is outright destabilized, while in other cases it ``jumps'' through the resonance instantaneously with no significant changes to the orbit, and we find that the probability for each outcome is about 20\%. Sometimes the final eccentricity is too low and Iapetus never encounters the secular resonance. A very common case is when the inclination of Iapetus increases or decreases slightly, which shifts the location of the secular resonance to substantially lower or higher eccentricities, respectively (Fig. \ref{loc}). Often we end up with an outcome where there is no $g_5$ secular resonance at all for $e > 0$, or it is shifted to high eccentricities and therefore will be ``missed" by Iapetus if its eccentricity is comparable to the present one. While this does not invalidate the idea that the resonance was crossed, it does make this process hard to model numerically. Additionally, the chaotic phase of the 5:1 MMR can last anything from a few to hundreds of Myr, making these integrations very computationally expensive given the uncertain outcome. Therefore, in order to complete a large number of simulations in reasonable time, we used a sped-up tidal evolution with $Q/k_2=500$ for Saturn. Figures \ref{sres42} and \ref{sres34} show two such integrations (made using {\sc ssimpl} with $\alpha=0.235$, case D in Fig. \ref{prec}). In Fig. \ref{sres42} Iapetus is captured into the secular resonance, while in Fig. \ref{sres34} it jumps through the resonance. Resonance capture happens only in about 10-20\% of our outcomes (which assume $e=0$ and the current inclination for Iapetus before the 5:1 MMR), making it somewhat unlikely, but not prohibitively so. We find that resonance ``jumps" and captures were about equally likely (with Iapetus not encountering the secular resonance in the rest of the cases). 

In order to verify this rate of resonant capture, we also performed a number of simulations using the realistic tidal properties of Saturn ($Q/k_2=5000$), but starting Titan immediately outside the 5:1 MMR with Iapetus, with Iapetus having its current or a somewhat higher eccentricity. While not as comprehensive as simulations which include the 5:1 resonance crossing, these runs should reflect the range of outcomes that are possible if Iapetus exits the MMR with $e$ and $i$ close to current values. This time we find that ``jumps'' through the resonance are an order of magnitude more common than captures. Apparently, the direction of resonance encounter we have here does not lead to capture under adiabatic conditions, and slower resonance encounters are ``worse'' for capture than faster ones. It is only because of the extremely slow libration period within the resonance that capture is possible in our simulations at all (i.e. even the realistic tidal evolution simulations are barely adiabatic). Therefore, while we cannot exclude the secular resonance capture in the aftermath of the Titan-Iapetus 5:1 MMR crossing, this is not a likely outcome.  
\begin{figure}
	\includegraphics[width=\columnwidth]{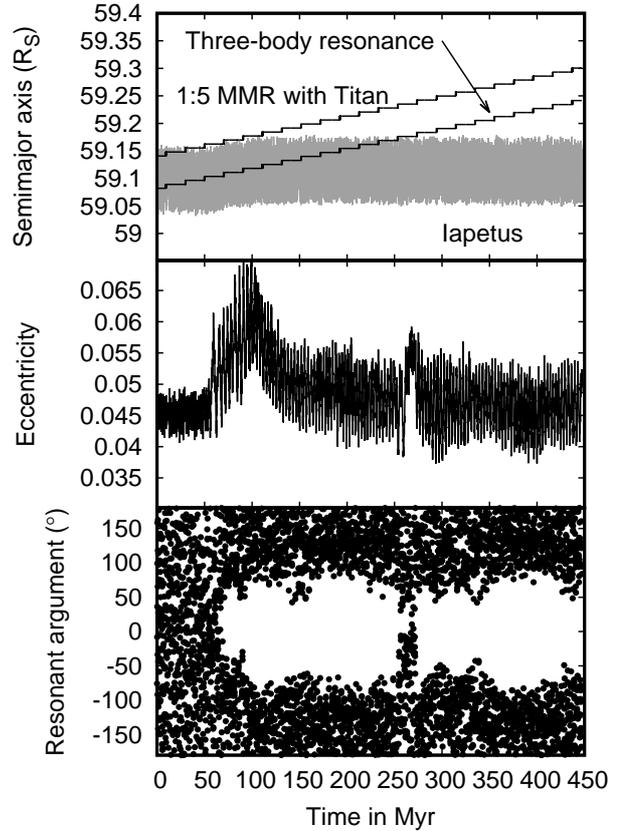}
    \caption{A simulation of the dynamics of Iapetus starting after the Titan-Iapetus 5:1 MMR using {\sc ssimpl}, with tidal $Q/k_2=5000$ for Saturn \citep{lai12}. The bottom panel plots the secular resonant argument $\varpi_I-\varpi_J+\Omega_I-\Omega_{eq}$. In this run, Iapetus is temporarily captured into a three-body resonance with partial argument $5\lambda - \lambda_T -\lambda_S$ between 50 and 100~Myr. During the three-body resonance, Iapetus enters the secular resonance and remains in it for most of the rest of the simulation.}
\label{sres52}
\end{figure}
Realistic-rate simulations of the 5:1 MMR aftermath uncovered some additional dynamical effects. Hundreds of Myr after Titan has crossed the main 5:1 MMR with Iapetus, new mean-motion-type resonances are observed in our simulation. These resonances seem to happen at Titan's semimajor axes that are an integer number of Saturn's mean motions away from the 5:1 resonance with Iapetus, meaning that they are three-body resonances involving Titan, Iapetus and the Sun. Fig. \ref{sres52} shows Iapetus encountering such a resonance, becoming temporarily captured in the three-body resonance (50-100 Myr). While Iapetus is in the three-body resonance, its eccentricity grows and the secular resonance is encountered. About 100 Myr into the simulation, Iapetus leaves the three-body resonance and remains in the secular resonance for the remainder of the simulation (with one short break). We are certain that the mean-motion part of the three-body resonance is $5\lambda - \lambda_T -\lambda_S$, but we were unable to find a complete librating argument, possibly indicating that the resonance is chaotic and what looks like temporary capture is constant shifting between different sub-resonances. In any case, we find that such three-body resonances are the major cause of both capture and escape from the secular resonance in realistic-rate simulations, greatly complicating the dynamics. We conclude that a capture into the  Iapetus-$g_5$ secular resonance following the Titan-Iapetus 5:1 MMR crossing is a possible but not a very likely outcome, and that additional work is needed to further examine the dynamics of this phase of the system's evolution.  

\begin{figure}
	\includegraphics[width=\columnwidth]{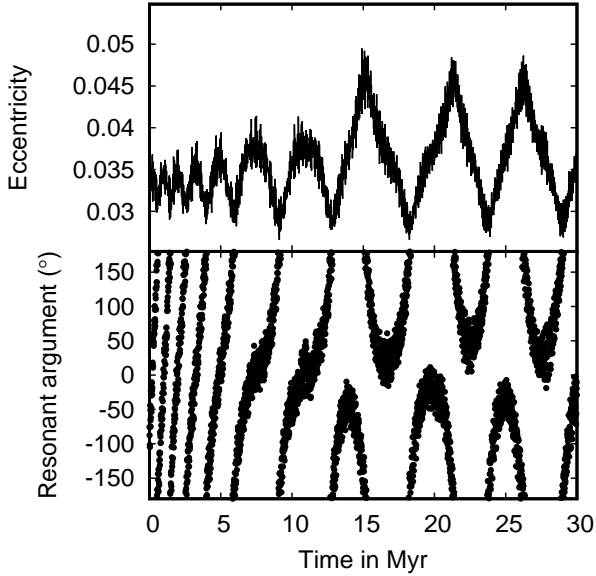}
    \caption{A simulation of the dynamics of Iapetus starting after the Titan-Iapetus 5:1 MMR using {\sc ssimpl}, with tidal $Q=100$ and $k_2=0.37$ for Saturn, implying orbital evolution dominated by Saturn's normal modes \citep{ful16}. The bottom panel plots the secular resonant argument $\varpi_I-\varpi_J+\Omega_I-\Omega_{eq}$. As in Fig. \ref{sres52}, we used the current inclination of Iapetus. In this run, the starting eccentricity of Iapetus is just right for it to be captured into the secular resonance. The capture is apparently non-adiabatic, and there is no noticeable evolution of Iapetus's orbit along the resonance.}
\label{fres72}
\end{figure}

Since the capture into the reasonance appears non-adiabatic, and simulations with examples shown in Figs. \ref{sres42}-\ref{sres52} indicate that faster migration of Titan improves the chances of resonance capture, we may want to reconsider some of our assumptions about Saturn's tidal response. While the $Q/k_2=5000$ for Titan we used so far is based on results of \citet{lai12}, theoretical predictions \citep{ful16} and some recent observational results \citep{lai17} suggest that the orbital evolution of Rhea and (possibly) Titan may be much faster than would be expected if Saturn's tidal $Q$ was the same for all satellites. Therefore we perfomed some additional integrations using $Q=100$ and $k_2=0.37$ for Saturn (V. Lainey, pers. comm.), with an example shown in Fig. \ref{fres72}. These integrations are similar to that shown in Fig. \ref{sres52} by starting with present $e$ and $i$ just after the Titan-Iapetus MMR. The dynamics of the 5:1 resonance crossing in the $Q \simeq 100$ regime is beyond the scope of this paper and is addressed by \citet{pol17} and \citet{pol18}. We find that the evolution is non-adiabatic as the evolution is fast relative to resonant librations, and capture is likely for a narrow range of initial eccentricities around $e=0.03$, but impossible for any other $e$. If Titan's orbit does indeed evolve this fast, then the Iapetus-$g_5$ secular resonance is an accidental side-effect of the stochastic 5:1 Titan-Iapetus MMR crossing. While this mechanism of resonance capture is promising and appears more straightforward than the one shown in Fig. \ref{sres52}, more work on the Titan-Iapetus 5:1 MMR is needed to properly evaluate the probabilities of either scenario. Note that a sustained rapid orbital evolution of Titan (equivalent to Saturn's tidal $Q \simeq 100$) would make the secular resonance only about 50~Myr old, which would naturally explain its existence despite its apparent dynamical fragility. 

\section{Summary}

This paper represent a first exploration of a previously unknown orbital resonance between Iapetus and the planetary system. Our conclusions can be summarized as follows:

1. Iapetus is currently in a secular resonance with an argument $\varpi-\Omega+\varpi_J-\Omega_{eq}$ librating around 180$^{\circ}$. The libration period is several Myr and the libration is likely to persists for several tens of Myr.

2. Longer-term stability of this resonance is tied to the precession of Saturn's spin axis, and more definite predictions need to wait for better determinations of Saturn's precession rate. Most allowable solutions for Saturn's pole precession lead to eventual breaking of the Iapetus-$g_5$ secular resonance, but some solutions preserve the secular resonance for at least 100 Myr.

3. We use analytical considerations to establish that the current occurence of the Iapetus-$g_5$ secular resonance is enabled by near-canceling of the precession term $\dot{\varpi} +\dot{\Omega}$, arising from several different secular and averaged periodic terms in the disturbing function. 

4. The Iapetus-$g_5$ secular resonance was almost certainly established more recently than the proposed 5:1 MMR crossing between Titan and Iapetus (500-50 Myr ago, depending on the Titan's unknown tidal evolution rate). While we find cases when the secular resonance was established in the aftermath of this MMR (with the more rapidly evolving Titan offering promissing results), we yet have to find a high-probability mechanism for establishing the secular resonance. More work is clearly needed to fully understand the rich dynamical history of Iapetus.

\section*{Acknowledgements}

Work by M{\'C} on this project was supported by NASA Outer Planets Research Program award NNX14AO38G. The authors thank Valery Lainey and William Polycarpe for very useful discussions. We thank the reviewer Doug Hamilton for alerting us to very relevant work by Callegari and Yokoyama.







\bsp	
\label{lastpage}
\end{document}